\documentstyle[12pt,epsf]{article}
\newcommand{\nub}{\overline{\nu}}

\newcommand{\sinttw}{\mbox{$\sin^2\theta_W$}}
\newcommand{\stw}{\mbox{$\sin^2\theta_W$}}

\newcommand{\rnu}{\mbox{$R^{\nu}$}}

\newcommand{\rmt}{\rm\textstyle}
\newcommand{\nunuc}{\mbox{$\nu$N}}
\def\be{\begin{equation}}
\def\ee{\end{equation}}
\def\bea{\begin{eqnarray}}
\def\eea{\end{eqnarray}}
\def\beas{\begin{eqnarray*}}
\def\eeas{\end{eqnarray*}}
\def\err#1#2#3  {{\it Erratum} {\bf#1} (#2) #3 }
\def\ib#1#2#3   {{\it ibid.} {\bf#1} (#2) #3 }
\def\ijmp#1#2#3 {{\em Int. J. Mod. Phys.} {\bf#1} (#2) #3 }
\def\jetp#1#2#3 {{\em JETP Lett.} {\bf#1} (#2) #3 }
\def\mpl#1#2#3  {{\em Mod. Phys. Lett.} {\bf#1} (#2) #3 }
\def\nat#1#2#3  {{\em Nature (London)} {\bf#1} (#2) #3 }
\def\nc#1#2#3   {{\em Nuovo Cim.} {\bf#1} (#2) #3 }
\def\nim#1#2#3  {{\em Nucl. Instr. Meth.} {\bf#1} (#2) #3 }
\def\np#1#2#3   {{\em Nucl. Phys.} {\bf#1} (#2) #3 }
\def\pcps#1#2#3 {{\em Proc. Cam. Phil. Soc.} {\bf#1} (#2) #3 }
\def\pl#1#2#3   {{\em Phys. Lett.} {\bf#1} (#2) #3 }
\def\prep#1#2#3 {{\em Phys. Rep.} {\bf#1} (#2) #3 }
\def\prev#1#2#3 {{\em Phys. Rev.} {\bf#1} (#2) #3 }
\def\prl#1#2#3  {{\em Phys. Rev. Lett.} {\bf#1} (#2) #3 }
\def\prs#1#2#3  {{\em Proc. Roy. Soc.} {\bf#1} (#2) #3 }
\def\ptp#1#2#3  {{\em Prog. Th. Phys.} {\bf#1} (#2) #3 }
\def\rmp#1#2#3  {{\em Rev. Mod. Phys.} {\bf#1} (#2) #3 }
\def\rpp#1#2#3  {{\em Rep. Prog. Phys.} {\bf#1} (#2) #3 }
\def\sjnp#1#2#3 {{\em Sov. J. Nucl. Phys.} {\bf#1} (#2) #3 }
\def\spj#1#2#3  {{\em Sov. Phys. JEPT} {\bf#1} (#2) #3 }
\def\zp#1#2#3   {{\em Zeit. Phys.} {\bf#1} (#2) #3 }
\begin{document}
\newcommand{\linespace}[1]{\protect\renewcommand{\baselinestretch}{#1}
  \footnotesize\normalsize}
\begin{flushright} 
FERMILAB Conf-96/227
\end{flushright}
\begin{center} 
\vspace*{3.5cm} 
UPDATED ELECTROWEAK MEASUREMENTS FROM NEUTRINO-NUCLEON
DEEPLY INELASTIC SCATTERING AT CCFR
\end{center} 
{\footnotesize
\begin{center}
\begin{sloppypar}
\noindent
        K.~S.~McFarland$^5$, C.~G.~Arroyo$^4$, P.~Auchincloss$^8$,  
        P.~de~Barbaro$^8$, A.~O.~Bazarko$^4$, R.~H.~Bernstein$^5$, 
        A.~Bodek$^8$, T.~Bolton$^6$,
        H.~Budd$^8$, J.~Conrad$^4$, R.~B.~Drucker$^7$,D.~A.~Harris$^8$,  
        R.~A.~Johnson$^3$, J.~H.~Kim$^4$, B.~J.~King$^4$, 
        T.~Kinnel$^9$, G.~Koizumi$^5$, S.~Koutsoliotas$^4$,
        M.~J.~Lamm$^5$, W.~C.~Lefmann$^1$, 
        W.~Marsh$^5$, C.~McNulty$^4$, 
        S.~R.~Mishra$^4$, D.~Naples$^5$, P.~Nienaber$^{10}$, 
        M.~Nussbaum$^3$, M.~J.~Oreglia$^2$, 
        L.~Perera$^3$, P.~Z.~Quintas$^4$, A.~Romosan$^4$,
        W.~K.~Sakumoto$^8$, B.~A.~Schumm$^2$,
        F.~J.~Sciulli$^4$, W.~G.~Seligman$^4$, M.~H.~Shaevitz$^4$, 
        W.~H.~Smith$^9$, P.~Spentzouris$^4$,
        R.~Steiner$^1$, E.~G.~Stern$^4$,
        M.~Vakili$^3$, U.~K.~Yang$^8$ 
\vspace{15pt} 

$^1$ Adelphi University, Garden City, NY 11530 
$^2$ University of Chicago, Chicago, IL 60637 \\
$^3$ University of Cincinnati, Cincinnati, OH 45221 
$^4$ Columbia University, New York, NY 10027 \\
$^5$ Fermi National Accelerator Laboratory, Batavia, IL 60510 
$^6$ Kansas State University, Manhattan, KS 66506 \\
$^7$ University of Oregon, Eugene, OR 97403 
$^8$ University of Rochester, Rochester, NY 14627 \\
$^9$ University of Wisconsin, Madison, WI 53706 
$^{10}$ Xavier University, Cincinnati, OH 45207 
\end{sloppypar}
\end{center}
}
\begin{center} 
Talk Presented by K.~S.~McFarland 
\end{center} 
\vspace{5cm} 
\begin{abstract} 
%\rightskip=1.5pc
%\leftskip=1.5pc
We report the results of a study of electroweak parameters from observations
of neutral-current $\nu$N deeply inelastic scattering in the CCFR detector 
at the FNAL Tevatron Quadrupole Triplet neutrino beam.  An improved
extraction of the weak mixing angle in the on-shell renormalization scheme,
incorporating additional data and with an improved technique for constraining
systematic errors, is presented.  Within the Standard Model, this result
constrains the W mass with a precision comparable to that from direct
measurements.  The result is also presented in a model-independent form,
as constraints on neutral-current quark-neutrino couplings, to facilitate
comparisons with theories outside the Standard Model.  Using this result,
limits on new four-fermion interactions, leptoquarks and neutrino oscillations
are presented.  Prospects for a successor experiment, NuTeV (FNAL-E815), 
are also presented.
\end{abstract} 

\pagebreak
%\normalsize
%\textwidth 17.cm

\section{Introduction}

In neutrino-nucleon (\nunuc) scattering, the ratio of neutral current
($Z$ exchange) to charged current ($W$ exchange) cross-sections
is related to the neutral current quark couplings by the Llewellyn-Smith
formula \cite{llewellyn}:
\bea
\rnu & \equiv & \frac{\sigma(\nu_{\mu}N\rightarrow\nu_{\mu}X)}
                 {\sigma(\nu_{\mu}N\rightarrow\mu^-X)} \\ 
& = & (g_L^2+rg_R^2),
\label{eqn:ls}
\eea 
where
\bea
r & \equiv & \frac{\sigma({\overline \nu}_{\mu}N\rightarrow\mu^+X)}
                {\sigma(\nu_{\mu}N\rightarrow\mu^-X)},  
\label{eqn:rdef} 
\eea 
and $g_{L,R}^2=u_{L,R}^2+d_{L,R}^2$, the isoscalar sum of the squared
left or right-handed quark couplings.  There are small corrections to this
relation from higher-twist effects, isovector components to the nuclear
target and electromagnetic radiative effects, and a substantial
correction from massive quark effects, such as scattering from the strange
or charm sea.  Because the neutral current quark couplings are functions
of the weak mixing angle, $\theta_W$, a measurement of \rnu\ can be used
to extract \stw.

Electroweak radiative corrections introduce significant $M_{{\rmt top}}$ 
and $M_{{\rmt Higgs}}$ dependences into these couplings.  However, in the
``on-shell'' (Sirlin) Renormalization scheme where
$\sinttw\ \equiv 1-\frac{M_W^2}{M_Z^2}$, the one-loop electroweak radiative
corrections to the quark couplings cancel approximately
in equation~\ref{eqn:ls} \cite{sirlin}.  
Therefore, if the quantity \rnu\ is measured and used to extract
a value of \stw, the result will be almost equal to
$1-\frac{M_W^2}{M_Z^2}$.  Given the very precise measurement of
the $Z$ mass from the LEP experiments, \nunuc\ scattering can, within
the Standard Model, provide a precise measurement of $M_W$ at energies
far below $W$ production threshold.

Outside the Standard Model, deviations between electroweak parameters
measured in \nunuc\ scattering and other processes are sensitive to
a host of new physics possibilities \cite{langacker}.  Possibilities
discussed in this paper include new four-fermion contact interactions at
high mass scales, leptoquarks and neutrino oscillations.  Discussed
elsewhere in these proceedings is the possibility that \nunuc\ scattering may
be sensitive to ``leptophobic'' $Z'$ bosons 
\cite{dibartolomeo}\cite{myrantings}.

\section{Experimental Technique}

The CCFR detector consists of an $18$~m long,
$690$~ton target calorimeter with a mean density of $4.2$~g/cm$^3$,
followed by an iron toroid spectrometer.  The target calorimeter
consists of 168 iron plates, $3$m~$\times$~$3$m~$\times$~$5.1$cm each. 
The active elements are liquid scintillation counters spaced
every two plates and drift chambers spaced every four plates.
There are a total of 84 scintillation counters and 42 drift chambers
in the target. The toroid spectrometer is not directly used in this
analysis.

The Tevatron Quadrupole Triplet neutrino beam is created by decays of
pions and kaons produced when $800$~GeV protons hit a production target.
A wide band of secondary energies is accepted by focusing magnets. 
The production target is located about $1.4$~km upstream of the
neutrino detector. The production target and focusing train are
followed by a $0.5$~km decay region. 
%The resulting
%neutrino energy spectra for $\nu_\mu$, $\overline{\nu}_\mu$, $\nu_e$,
%and $\overline{\nu}_e$ at the detector are shown in Figure
%\ref{fig:enu}.  
The beam is predominantly muon neutrinos and anti-neutrinos, but
contains a small fraction of electron neutrinos ($2.3$\%)  
and a negligible fraction of tau neutrinos (less than $10^{-5}$) which
result primarily from $D_s$ decay.

Neutrinos are observed in the target calorimeter {\it via} their neutral
current and charged current interactions.  $\nu_\mu$ charged current events
are characterized by the presence of a muon in the final
state which deposits energy in a large number of consecutive scintillation
counters as it travels through the calorimeter.  Neutral current
events have no muon and deposit energy over a range of counters
typical of a hadronic shower (5 to 20 counters).  Accordingly,
we define ``short'' events as those which deposit energy over an
interval of 30 or fewer scintillation counters.
The ratio $R_{30}$ is defined to be the number of
short events divided by the number of long events \cite{wma}.  

We define $E_{cal}$ as the energy deposited in the calorimeter in the
first twenty counters following the event vertex.  
Events were selected using a calorimeter trigger fully sensitive for
$E_{cal}$ above $20$~GeV, and only events with $E_{cal}$ above $30$~GeV
were used in the analysis.  To ensure event containment, the fiducial
volume of the detector is limited to a central cylindrical region 30''
in radius and excludes events which began in the first 6 counters or
the last 34 counters of the detector.  The resulting data sample
consisted of about 660,000 events.

A detailed Monte Carlo was used to determine electroweak parameters 
from the measured $R_{30}$.  The only undetermined inputs to this Monte 
Carlo were the neutral current quark couplings which were then varied until
the Monte Carlo predicted an $R_{30}$ which agreed with that observed
in the data.  For the extraction of \stw, the couplings in the Monte Carlo
were fixed to their Standard Model predictions as functions of \stw\ which
was then varied as the only free parameter.  The Monte Carlo included 
detector response and beam simulations,
as well as a detailed cross-section model which included electromagnetic
radiative corrections, isovector target corrections, heavy quark production
and seas, the longitudinal cross-section and lepton mass effects.

\begin{table}[bt]
\begin{minipage}[t]{8.0cm}
\begin{tabular}{|r|c|}
\hline
  SOURCE OF UNCERTAINTY & $\delta\sin^2\theta_W$ \\
\hline
 { data statistics } & {\fbox{\bf 0.0021}}  \\
 { Monte Carlo statistics } & { 0.0005} \\ \hline
 { TOTAL STATISTICS \hfill } & { 0.0021} \\ \hline\hline
  { Charm Production } &   \\
  { ($m_c=1.31\pm0.24$~GeV)} & {\fbox{\bf 0.0029}}  \\
  { Charm Sea} & {\fbox{\bf 0.0014}} \\
  { Longitudinal Cross-Section} & { 0.0008}  \\
   { Higher Twist} & { 0.0005} \\
   { Non-Isoscalar Target} & { 0.0003} \\
  { Strange Sea} & { 0.0003}  \\
   { Structure Functions} & { 0.0002} \\ 
  { Rad. Corrections} & { 0.0001} \\ \hline
  { TOTAL PHYSICS MODEL \hfill } & { 0.0034} \\  \hline  
\end{tabular}
\end{minipage}\hspace{0.5cm}
\begin{minipage}[t]{8.0cm}
\begin{tabular}{|r|c|}
\hline
  SOURCE OF UNCERTAINTY & $\delta\sin^2\theta_W$ \\
\hline
  { $\nu _e$ flux ($4.2\%$)} & {\fbox{\bf 0.0022}} \\
  { Transverse Vertex} & { 0.0008} \\
  { Energy Measurement  \hfill } & \\
  { Muon Energy Loss in Shower} & { 0.0006} \\
  { Absolute Energy Scale ($1\%$)} & { 0.0004} \\
  { Hadron Energy Scale ($0.6\%$)} & { 0.0002} \\
  { Event Length     \hfill } & \\
  { Hadron Shower Length} & { 0.0006} \\
  { Vertex Determination} & { 0.0008} \\
  { Counter Efficiency and Noise} & { 0.0006} \\
  { Dimuon Production} & { 0.0003} \\
  \hline
  { TOTAL EXP. SYST. \hfill } & { 0.0027} \\ \hline\hline
  { TOTAL UNCERTAINTY \hfill } & { 0.0048} \\  \hline
\end{tabular}
\end{minipage}
\caption{Uncertainties in the {\em preliminary} extraction of \stw\ from
the CCFR data}
\label{errors}
\end{table}

There are three major uncertainties in the comparison of $R_{30}$
from the Monte Carlo to the data: the statistical error in
the data, the uncertainty in the effective charm quark mass for
charged current charm production, the uncertainty in the incident flux of
$\nu_e$'s on the detector.  Other sources of systematic
uncertainty  were also investigated \cite{wma}.   Table~\ref{errors} 
shows the effect of the uncertainties on the determination of \stw.

The charm mass error comes from the uncertainty in modeling the
turn-on of the charm quark production cross section.  The Monte Carlo
uses a slow-rescaling model with the parameters extracted using events
with two oppositely charged muons in this experiment \cite{baz}.  This
error dominates the calculation of $R_{30}$ at low $E_\nu$ (and low
$E_{cal}$) where the threshold suppression is greatest.  The $\nu_e$
flux uncertainty has a large effect on $R_{30}$ because almost all
charged current $\nu_e$ events are short events.  Therefore, the
relatively small ($4.2\%$ \cite{wma}) fractional uncertainty in the
$\nu_e$ flux is a large effect, particularly at high $E_{cal}$ since
most $\nu_e$ charged current interactions deposit the full incident
neutrino energy into the calorimeter.  This $4.2\%$ is dominated by a
$20\%$ production uncertainty in the $K_L$ content of the secondary
beam which produces $16\%$ of the $\nu_e$ flux.  The bulk of the
$\nu_e$ flux comes from $K^{\pm}_{e3}$ decays, which are
well-constrained by the observed $\nu_\mu$ spectrum from $K^{\pm}_{\mu
2}$ decays \cite{wma}.

\section{Results}

\begin{figure}[bt]
\epsfxsize=\textwidth
\epsfbox{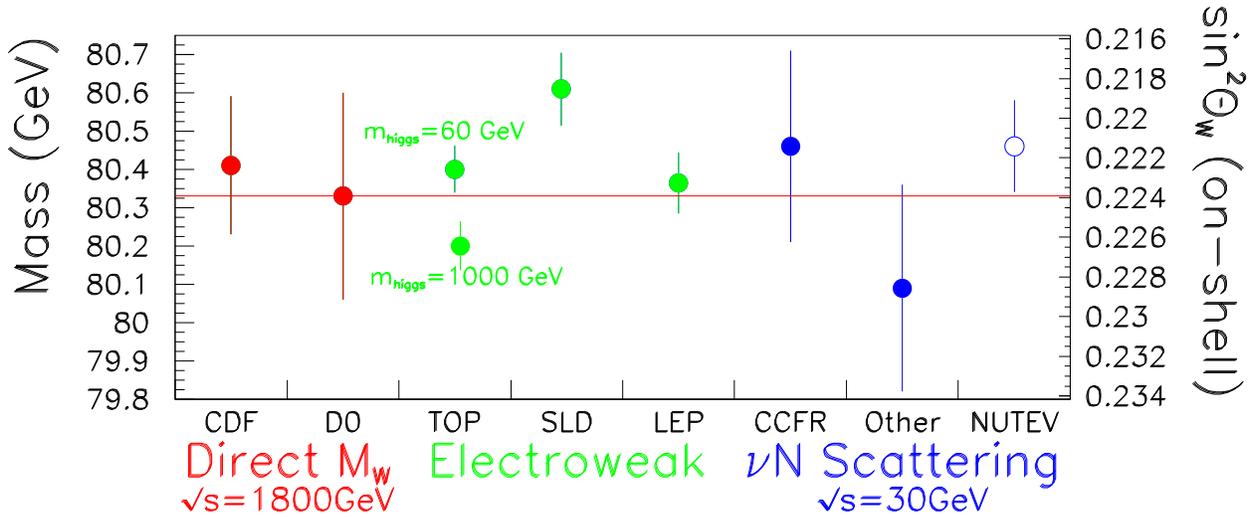}
\caption{A comparison of different precision electroweak measurements,
shown as a predicted $W$ mass within the Standard Model}
\label{fig:wmass}
\end{figure}

CCFR has updated its previously published result \cite{wma} with the
addition of more data and an improved analysis of systematic errors.
The new {\em preliminary} result from CCFR for the weak mixing angle
in the on-shell renormalization scheme is:
\be
\sin^2\theta_W=0.2213\pm0.0021({\rmt stat})
\pm0.0027({\rmt syst})\pm0.0034({\rmt model}).
\label{eqn:wmares} 
\ee
The additional uncertainty on this on-shell \stw\ from $m_{\rmt
top}=174\pm10$~GeV due to the one-loop electroweak radiative
corrections is $\pm0.0003$. Within the Standard Model, this
corresponds to a $W$ mass of $80.46\pm0.25$~ GeV.  It is possible to
combine the world's \nunuc\ scattering data on isoscalar targets and
obtain an average.  However, because of the large charm production
systematic which is common to all experiments, there is not much
improvement.  Combining the five most precise experiments,
$\sin^2\theta_W=0.2261\pm0.0040$, with a $\chi^2/$DOF of $5.33/4$.
Shown in Figure~\ref{fig:wmass} is the good agreement of equivalent
$W$ mass measurements from \nunuc\ experiments, direct measurements
from the Tevatron, and derivations from $Z^0$ observables and $m_{\rmt
top}$.

\begin{figure}[bt]
\epsfxsize=\textwidth
\epsfbox{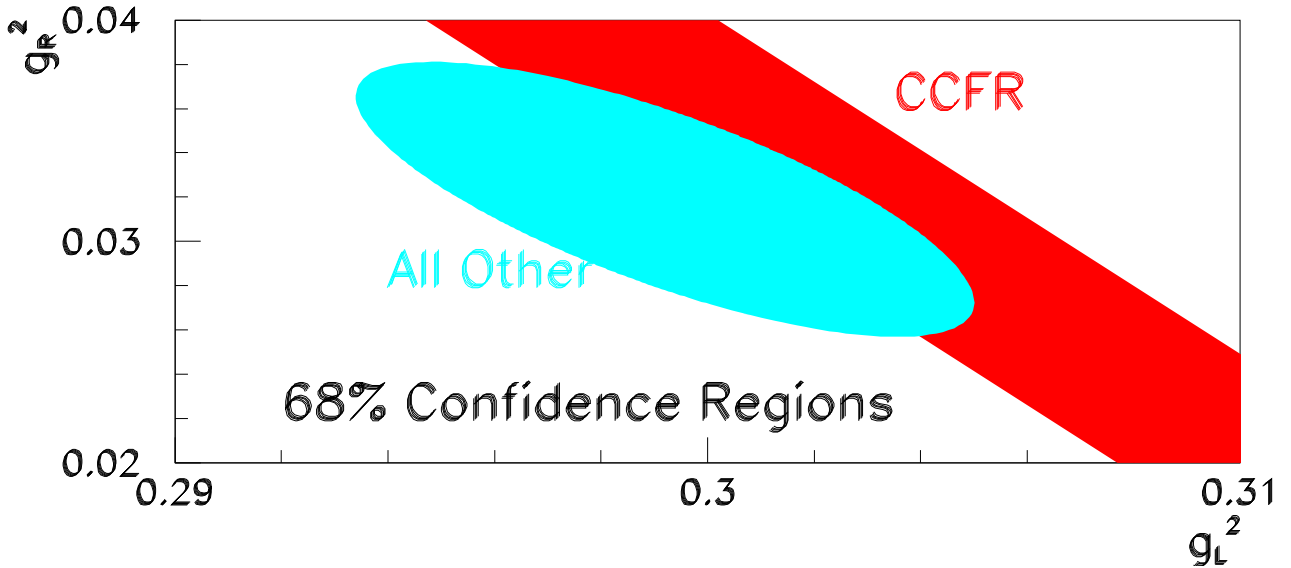}
\caption{One-sigma constraints on the isoscalar neutral current quark 
  couplings, $g_L^2$ and $g_R^2$, from this result and other neutrino data.}
\label{fig:couplings}
\end{figure}

To facilitate comparisons with extensions to the Standard Model, this
result can also be expressed as a model-independent constraint on the
neutral-current quark couplings.  The {\em preliminary} CCFR constraint
is
\be
\kappa=0.5629\pm0.0048=1.7266g_L^2+1.1198g_R^2
-0.1008\delta_L^2-0.0865\delta_R^2
\label{eqn:coupcon}
\ee
where $(\stackrel{g}{\delta})_{L,R}^2=u_{L,R}^2(\pm) d_{L,R}^2$.
The Standard Model prediction is $\kappa=0.5623\pm0.0016$ for the measured
values of $m_Z$, $m_{\rmt top}$, $m_W$.
Figure~\ref{fig:couplings} shows this result compared with a fit
to other neutrino data \cite{foglihaidt}.

\section{Constaints on New Physics}
The following sections make use of the value of $\kappa$ given in
equation~\ref{eqn:coupcon} and its Standard Model value to set limits on new physics
possibilities.

\subsection{Compositeness Scales}

One can postulate a four-fermion interaction between two neutrinos and
two quarks, and add a term to the interaction Lagrangian of the form
$-{\cal L}=\pm(4\pi/\lambda^{\pm}_{LL})\overline{l}_{\mu L}\gamma^\nu
l_{\mu L}\overline{q}_L\gamma_\nu q_L$.  This interaction will shift the 
predicted values for the neutral current quark couplings, and thus
the \nunuc\ data can limit the allowed range of $\lambda^{\pm}_{LL}$.
From the preliminary CCFR result, at 95\% confidence, 
$\lambda^{+}_{LL}>3.8$~TeV or $\lambda^{-}_{LL}>3.5$~TeV.

\subsection{Leptoquarks}

The model used for this search is an SU(5)-inspired model \cite{langacker}.
If there are no leptoquark-induced flavor-changing neutral currents and
if the left-handed coupling of the leptoquark ($\eta_L$) is much larger 
than its right-handed coupling, \nunuc\ is one of the most sensitive
probes.  From the preliminary CCFR result, at 95\% confidence,
$M_L/\left| \eta_L\right| > 0.8$~TeV.

\subsection{Neutrino Oscillations}

\begin{figure}[bt]
\epsfxsize=\textwidth
\epsfbox{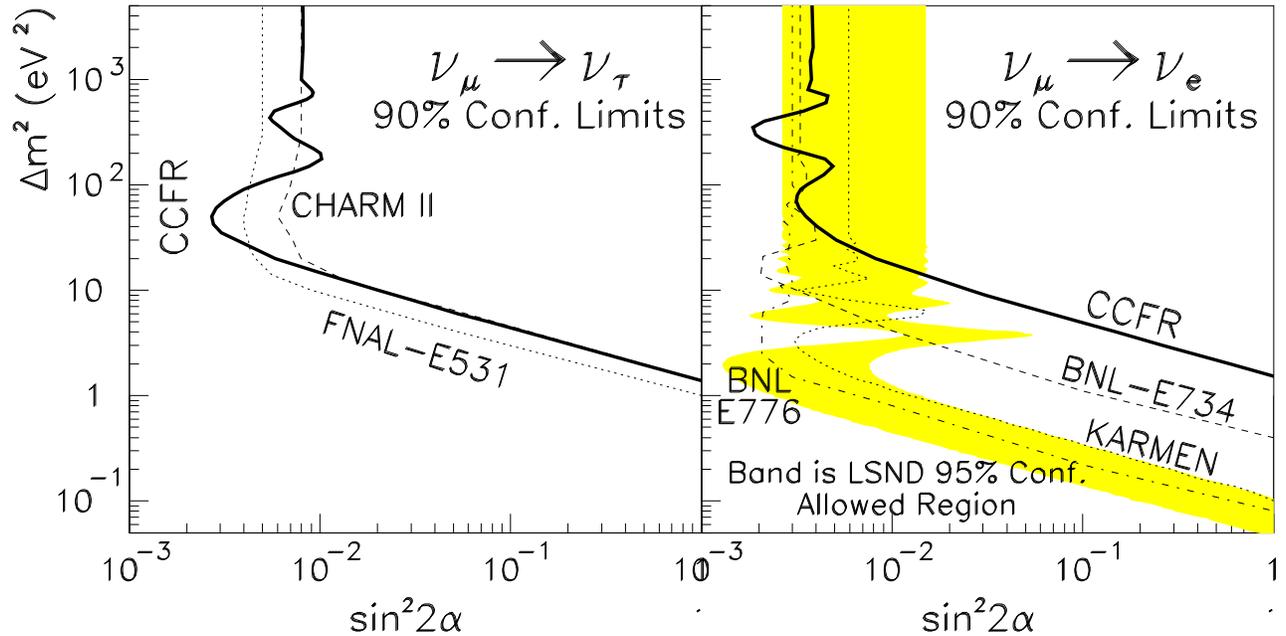}
\caption{90\% confidence limits on $\nu_\mu\to\nu_{\tau,e}$ oscillation 
from the CCFR electroweak measurement compared with othger experiments}
\label{fig:osc}
\end{figure}

Neutrino oscillations, if present, would also affect the measured neutral
current quark couplings.  This is because charged-current events are selected
by the presence of a muon in the final state, and clearly if muon neutrinos
oscillate to either electron or tau neutrinos, they are less likely to
produce final-state muons in their charged current interactions.  Details
of this analysis can be found elsewhere \cite{oscpaper}; the limits obtained
are shown in Figure~\ref{fig:osc}. 

\section{Conclusions}

Even in the era of high-luminosity colliders that produce copious
on-shell $W$ and $Z$ bosons, neutrino-nucleon deeply inelastic
scattering remains an interesting system in which to pursue
measurements of electroweak parameters.  Within the Standard Model,
the CCFR measurement of \stw\ from \nunuc\ scattering provides a 
measurement of the $W$ mass with comparable precision to current 
measurements at the Tevatron.  Outside the Standard Model, this measurement 
is sensitive to new physics at the TeV scale and to
neutrino oscillations.

The NuTeV experiment at Fermilab will continue to improve the
precision of measurements of \nunuc\ scattering.  A new beamline, the
Sign-Selected Quadrupole Train (SSQT), will run from 1996-1998 and
provide separate high-intensity neutrino and anti-neutrino beams,
This will allow separate measurements of the neutrino and
anti-neutrino neutral current cross-sections.  The difference,
$\sigma^\nu_{NC}-\sigma^{\nub}_{NC}$, is insensitive to sea quark
distributions, and will allow a measurement of \stw\ with model errors
reduced by a factor of 3.  The SSQT also produces almost no electron
neutrinos from $K_L$ decays, thus removing the dominant source of
experimental uncertainty in the CCFR measurement.
NuTeV projects a precision of $\pm0.0019$ in its measurement of 
\stw\ which corresponds within the Standard Model to a
 $W$ mass precision of $100$~MeV.

\end{document}